\pdfoutput=1

\documentclass[twocolumn, times]{aastex63} 

\usepackage{natbib}
\usepackage{amsmath}
\usepackage{amssymb}
\usepackage{subfigure}
\usepackage{url}
\usepackage{CJK}
\usepackage{longtable}

\renewcommand{\vec}[1]{ {\mathbf #1} }

\newcommand{\Fig}{{Figure}}
\newcommand{\Figs}{{Figures}}

\newcommand{\SDO}{{\it SDO}}

\shorttitle{Test data-driven model}
\shortauthors{Jiang et al.}

\begin{document}

\title{Testing a Data-driven Active Region Evolution Model with
  Boundary Data at Different Heights from a Solar Magnetic Flux
  Emergence Simulation}

\correspondingauthor{Chaowei Jiang} \email{chaowei@hit.edu.cn}

\author[0000-0001-8605-2159]{Chaowei Jiang}
\affiliation{Institute of Space Science and Applied Technology, Harbin
  Institute of Technology, Shenzhen 518055, China}

\author[0000-0002-1276-2403]{Shin Toriumi} \affiliation{Institute of Space and
  Astronautical Science, Japan Aerospace Exploration Agency, 3-1-1
  Yoshinodai, Chuo-ku, Sagamihara, Kanagawa 252-5210, Japan}

\begin{abstract}
  A data-driven active region evolution (DARE) model has been
  developed to study the complex structures and dynamics of solar
  coronal magnetic fields. The model is configured with typical
  coronal environment of tenuous gas governed by strong magnetic
  field, and thus its lower boundary is set at the base of the corona,
  but driven by magnetic fields observed in the photosphere. A previous
  assessment of the model using data from a flux emergence simulation
  (FES) showed that the DARE failed to reproduce the coronal magnetic
  field in the FES, which is attributed to the fact that the
  photospheric data in the FES has a very strong Lorentz force
  and therefore spurious flows are generated in the DARE model. Here
  we further test the DARE by using three sets of data from the FES
  sliced at incremental heights, which correspond to the photosphere,
  the chromosphere and the base of the corona. It is found that the
  key difference in the three sets of data is the extent of the
  Lorentz force, which makes the data-driven model perform very
  differently. At the two higher levels above the photosphere, the
  Lorentz force decreases substantially, and the DARE model attains
  results in much better agreement with the FES, confirming that the
  Lorentz force in the boundary data is a key issue affecting the
  results of the DARE model. However, unlike the FES data, the
  photospheric field from {\SDO}/HMI observations has recently been
  found to be very close to force-free. Therefore, we suggest that it
  is still reasonable to use the photospheric magnetic field as
  approximation of the field at the coronal base to drive the DARE model.
\end{abstract}


\section{Introduction}
\label{sec:intro}

On the solar surface, i.e., the photosphere, magnetic fields are seen to change
continuously; magnetic flux emergence brings new flux from the solar
interior into the atmosphere, and meanwhile the flux is advected and dispersed by surface motions
such as granulation, differential rotation and meridional circulation.
Although we are still not able to measure directly the three-dimensional (3D) magnetic field in
the atmosphere, in particular, the solar corona, it is believed that
the coronal field evolves in response to (or driven by) the changing
of the photospheric field. Consequently, complex dynamics occur ubiquitously in the corona,
including the interaction of newly emerging field with the pre-existing one, twisting and
shearing of the magnetic arcade fields, magnetic reconnection, and magnetic explosions
 which are manifested as flares and coronal mass
ejections.

In macroscopic scale, evolution of the magnetic field in the solar
atmosphere is governed by the magnetohydrodynamic (MHD)
equations. With magnetic field in the photosphere measured routinely,
data-driven models using the photospheric magnetograms as boundary
conditions have been proposed in recent years to study the dynamic
evolution of solar coronal magnetic field~\citep[e.g.,][]{Wu2006,
  FengX2012, Cheung2012, Jiang2016NC, Leake2017, Inoue2018,
  Hayashi2019, GuoY2019, Pomoell2019}. Due to the limited constraint
from observation, data-driven models are developed with very different
settings from each other. For simplicity, some used the
magneto-frictional model~\citep{Cheung2012, Pomoell2019}, in which the
Lorentz force is balanced by a fictional plasma friction force, and
the magnetic field evolves mainly in a quasi-static way. Some used the
so-called zero-$\beta$ model~\citep{Inoue2018, GuoY2019}, in which the
gas pressure and gravity are neglected. It is more realistic to solve
the full MHD equations to deal with the nonlinear interaction of the
magnetic field with the plasma.

To solve the MHD equations, one needs to specify eight variables
(namely plasma density, temperature, and three components of velocity
and magnetic field, respectively) in a self-consistent way at the
lower boundary as well as in the initial conditions. However, due to
the limited observations, they are under-specified. Thus, different
choices of initial and boundary conditions have been adopted. For the
initial conditions, the background atmosphere was set as a uniformly
distributed gas~\citep[e.g.,][]{Hayashi2019}, an isothermal gas
stratified by solar gravity~\citep[e.g.,][]{Jiang2016NC}, or a more
realistic, highly stratified one with a temperature profile of
different levels representing the photosphere, chromosphere and
corona~\citep[e.g.,][]{Jiang2016F, GuoY2019}. The initial settings
depend mainly on the interest of study; for instance, if one focuses
on the coronal magnetic field, an isothermal gas with typical coronal
temperature (e.g., $\sim 10^6$~K) and number density (e.g.,
$\sim 10^9$~cm$^{-3}$) would be sufficient. However, the choice of
typical plasma $\beta$ and Alfv{\'e}n speed in the calculation is more
relevant since in the numerical models it is the dimensionless
parameters that matter. So, if one simulates the corona using the full
MHD model, the plasma $\beta$ should be typically much less than
unity, and the Alfv{\'e}n speed should be much larger than the sound
speed.

Even using the MHD model with typical coronal plasma settings, it is
still problematic to specify the lower boundary conditions by using
observed data in the photosphere, since the bottom surface in the
model is assumed to be the coronal base rather than the photosphere.
Certainly, the physical behavior of the photosphere is distinct from
that at the coronal height. For example, the key parameter, the plasma
$\beta$ in the photosphere is of order unity, and the
plasma controls evolution of the magnetic fields. The Alfv{\'e}n speed
in the photosphere is of the order of several km~s~$^{-1}$, much
smaller than the coronal values. Thus, in principle, it is required to
use a highly stratified atmosphere model including all the layers from
the photosphere to the corona. For example, the simplest form of such
stratified atmosphere has been extensively used in MHD simulations of
magnetic flux tube emergence from below the
photosphere~\citep[e.g.,][]{Fan2001, archontis_emergence_2004,
  Toriumi2010}, which includes a thin isothermal layer representing
the photospheric/chromospheric layer (with temperature of about
$5000$~K), a narrow transition layer with temperature sharply rising
by about two orders of magnitude, and another isothermal layer for the
corona. But such settings demand a considerably large amount of
computing resources, mainly to resolve the small scale heights in the
photosphere and the transition layer, in which the gas density varies
by more than eight orders of magnitude within a few megameters. Thus,
most of the flux emergence simulations (FESs) currently available only
deal with relatively small spatial scales of a few tens of megameters
in full three dimensions and short time durations of, typically,
hours. The burden of computational resources would be too much if one
wants to simulate the long-term (e.g., days) evolution of
active-region (AR) size (e.g., hundreds of megameters), which is the
main purpose of the data-driven models.

An alternative way to circumvent such technical obstacles is to
consider the photospheric data as a reasonable approximation of the
corresponding variables at the coronal base. This is what has been
done in static extrapolation of coronal magnetic field from the
photospheric magnetograms, for instance, the most frequently-used
nonlinear force-free field extrapolations from vector
magnetograms~\citep[e.g.,][]{Wiegelmann2004,Valori2007,
  Wiegelmann2012solar, Jiang2013NLFFF,Inoue2014}. At relatively large scales,
e.g., size of a typical AR, it is generally accepted to approximate the magnetic field at the coronal base by the
photospheric field, considering that the large-scale AR field with
typical scales of hundreds of megameters do not change much
within the thin layer of several megameters from the photosphere,
where it is measured, to the coronal base. For instance,
\citet{Jiang2016NC} developed the first full MHD model for data-driven
active region evolution (DARE) by using time-sequence of vector
magnetograms obtained from \SDO/HMI as boundary conditions at the
coronal base. They managed to simulate the coronal magnetic field
evolution in a flux-emerging AR leading to an eruption and attained a
reasonable agreement with \SDO/AIA observations. Further applications
of this DARE model, e.g., in \citet{Jiang2016ApJ} and \citet{HeW2020},
reveals the formation and reconnection of current sheet and magnetic
flux rope involved in major solar flares. See \citet{Toriumi2019} for
further review on DARE models of flare-productive active regions.

Nevertheless, such approximation of magnetic field at the coronal base
by the photospheric field should be taken with cautions. For instance,
smoothing is required because magnetic structures are broadened from
the photosphere to the corona due to the expansion of the
fields. \citet{Yamamoto2012} showed that the field measured at the
chromosphere is best correlated with the simultaneously observed
photospheric field that is smoothed by a Gaussian window of about 2
arcsecs~\citep[see also][]{Kawabata2020}.  A more critical issue is the Lorentz force contained in the
photospheric field. Since the plasma $\beta$ is large, it is thought
that the photospheric field has a strong Lorentz force to balance the
gas pressure gradients. For instance, in classical models of solar
magnetic flux emergence, the toroidal flux generated in the tachocline
can rise through the convection zone with the aid of magnetic buoyancy force.
However, at the photosphere, it is trapped because of the strongly sub-adiabatic stratification there,
and its subsequent rising into the atmosphere must resort to
the Parker instability~\citep{Parker1955,
  shibata_nonlinear_1989, archontis_emergence_2004}, which is a kind of
magnetic Rayleigh-Taylor instability. To trigger the Parker instability, the
magnetic flux piles up shallowly below the photosphere, until
the Lorentz force is built up as large as the gas pressure gradient to hold up an extra amount of mass
against gravity, and further ascent of the magnetic flux depends mainly
on liberation of the gravity potential energy of this extra amount of
mass~\citep{newcomb_convective_1961, Acheson1979,
  archontis_emergence_2004}. This process has been termed
``two-step emergence''~\citep[e.g.,][]{Toriumi2010}, and it naturally results in a strongly
non-force-free photosphere, which has been indeed shown in typical
numerical simulations of idealized flux tube emergence from below the photosphere
into the corona.

Thus, if the photospheric magnetic field is introduced in the
data-driven MHD model that uses typical settings of atmosphere in the
corona, such a strong Lorentz force cannot be balanced by the tenuous
plasma. It can induce largely spurious plasma motions, which in turn
amplify the magnetic field by the magnetic induction equation, and
could make the evolution runaway. This is indeed observed in a recent
joint comparative study of 4 different data-driven models~\citep{Cheung2012, Jiang2016NC,
  GuoY2019, Hayashi2019}
using the photospheric magnetograms produced in a FES as input to their bottom boundaries~\citep{Toriumi2020}. It
was found in~\citet{Toriumi2020} that, although all data-driven
models successfully reproduced a flux rope structure, the quantitative
discrepancies are very large, and this is attributed mainly to the highly
non-force-free input photospheric field and to the different settings of
the background atmosphere in the models. Especially, the simulated magnetic flux rope
in \citet{Jiang2016NC}'s DARE model, which uses typical settings of tenuous
atmosphere in the corona (i.e., low plasma $\beta$ and high
Alfv{\'e}n speed), exhibits a significantly
larger size and stronger magnetic twist than those in the
FES as well as the other data-driven MHD models that use
much denser plasma at the lower
boundary~\citep[e.g.,][]{ Hayashi2019, GuoY2019}.  Moreover, the
magnetic energy and helicity produced by the DARE model are approximately
ten times larger than their original values in the FES. By checking the simulation
data near the lower boundary, \citet{Toriumi2020} showed clearly such a big discrepancy
arises from the fact that the strong Lorentz force (and torque) in the photospheric field in the FES,
originally balanced by the photospheric plasma gravity and pressure,
exerts directly on the tenuous plasma in the DARE model.  As a result, the
plasma is pushed upward and rotated strongly there, making the magnetic field expands
quickly and twisted freely, which finally leads to an over-amplification of magnetic energy
and helicity in the corona.

However, such a strong runaway expansion of the coronal field as well as
its unreasonable level of energy and helicity are not seen in
DARE simulations by using the observed magnetograms as the
driving boundary conditions~\citep[e.g.,][]{Jiang2016NC, Jiang2016ApJ,
  HeW2020}. That is, the result of DARE simulation driven by the FES output data behaves quite distinctly from that driven by actual observational data; the former shows an unreasonable runaway expansion while the latter does not. As mentioned before, the runaway expansion results from strong Lorentz force and torque resident in the boundary data from the FES. This led us to a hypothesis that the actual photospheric field
is close to force free, unlike the simulated one in the flux emergence
model. Very recently, \citet{DuanA2020} statistically surveyed how large
Lorentz force is contained in the photospheric magnetic fields observed
by \SDO/HMI. They computed the normalized Lorentz forces and torques for
emerging ARs using a substantially large
sample of vector magnetograms (a total number of 3536 set of vector
magnetograms from 51 ARs covering the time from 2010 to 2019),
and concluded that the photospheric field is
very close to a force-free state, since they have a rather small Lorentz force and
torque on average with normalized values of $\sim 0.1$, which is clearly not consistent with
theories as well as idealized simulations of flux emergence.
Furthermore, such a result appears to be not influenced by the sizes of
emerging ARs, the emergence rate, or the non-potentiality of the emerging field.

As mentioned before, if a highly forced bottom boundary data is introduced to
the DARE model, it fails to reproduce the coronal magnetic field. On
the other hand, the observed photospheric field is actually close to
force-free. So, the purpose of this paper is to confirm that with a
more force-free boundary field, the DARE model can reproduce the
coronal magnetic field that is in better agreement with the FES.
Although the magnetic field in the photospheric level in the FES is far
from force-free, it relaxes quickly to an approximately force-free
state when the field emerges into the atmosphere by a few scale
heights above the photosphere, because of the fast decreasing of the
plasma density and pressure with height~\citep{Fan2009}. Thus, to
obtain such a more force-free boundary field from the FES model, we
use the data sliced at two more heights above $z=0$ (corresponding to
the height of photosphere in the FES model), which is used to mimic
the data at the chromosphere and coronal base.

The paper is organized as follows. In Section~\ref{sec:model}, we give
a brief description of the FES and DARE models, and the Lorentz force
and torque in the data sets from FES that are used to drive the DARE
model. The results of using different boundary data for the DARE are
compared in Section~\ref{sec:res}. Then we conclude in
Section~\ref{sec:con}.

\begin{figure*}
  \centering
  \includegraphics[width=\textwidth]{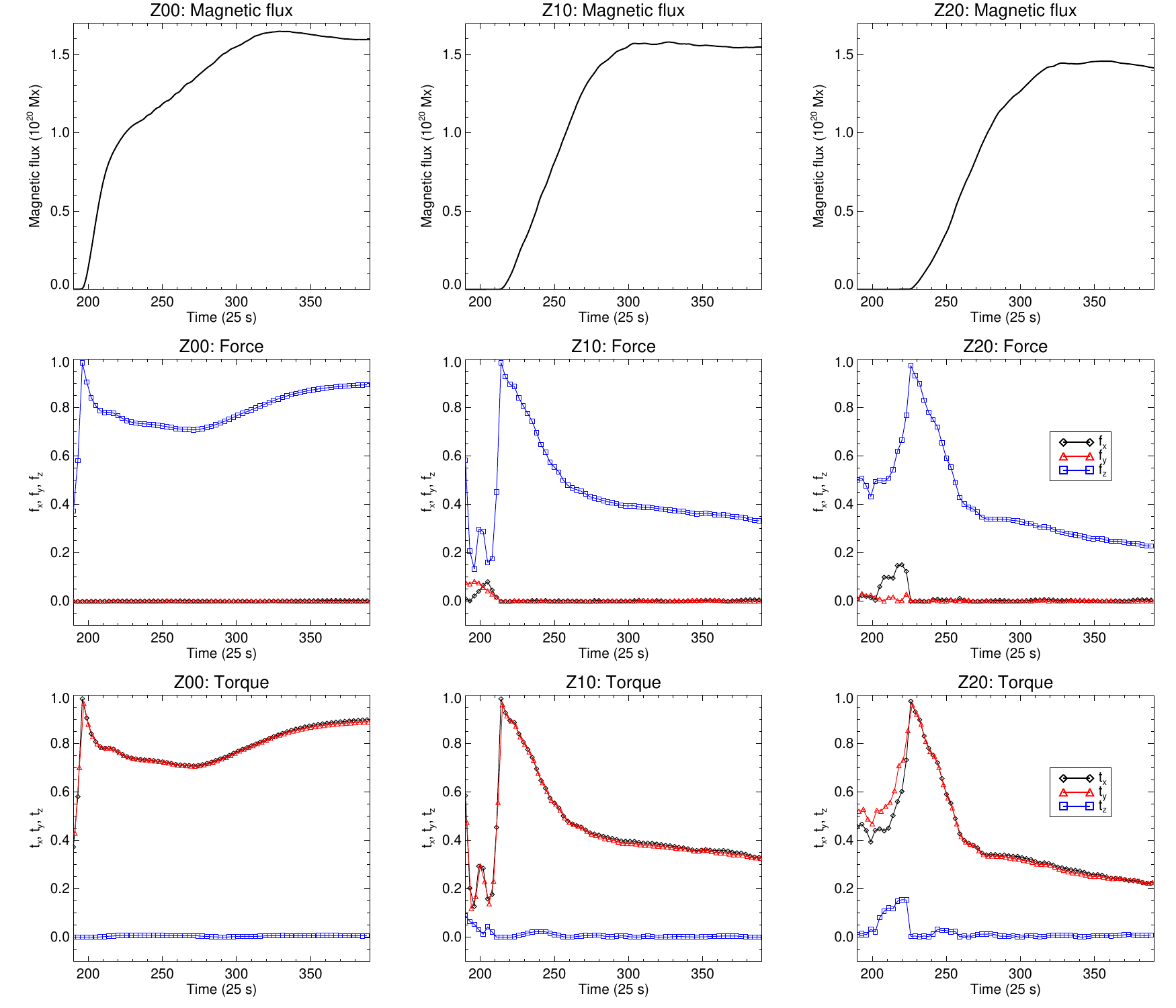}
  \caption{Magnetic flux (top), normalized Lorentz force (middle) and
    torque (bottom) in the FES data. From left to right are results
    for the \texttt{Z00}, \texttt{Z10}, and \texttt{Z20} slices,
    respectively}
  \label{residual_force}
\end{figure*}

\begin{figure*}
  \centering
  \includegraphics[width=\textwidth]{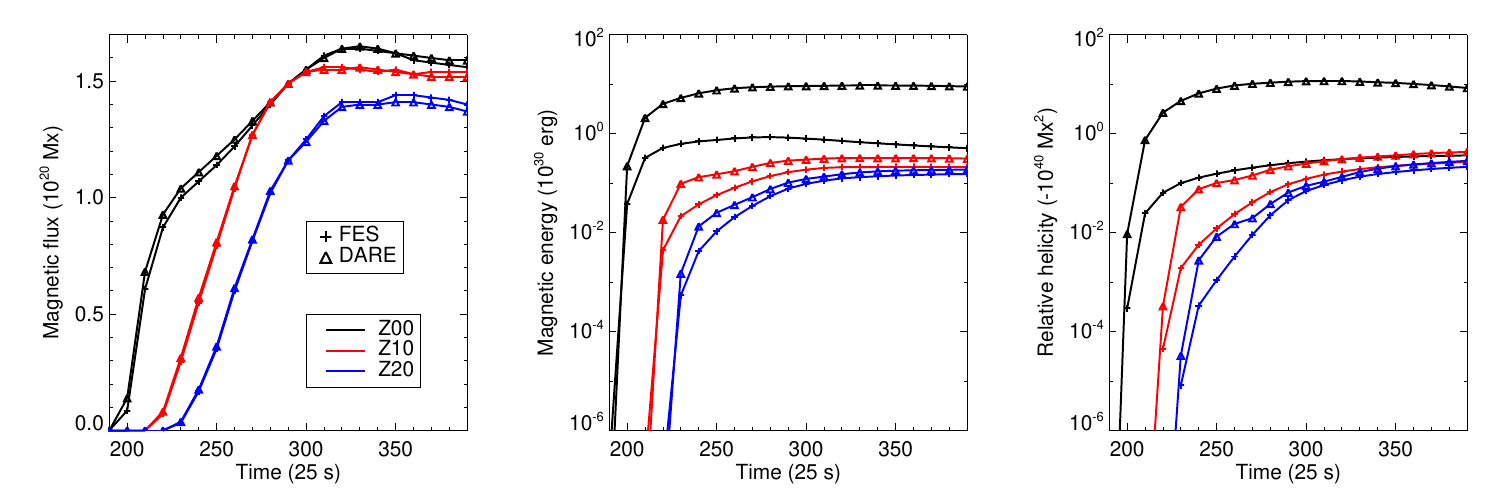}
  \caption{Comparison of magnetic flux, energy and relative helicity
    in the three DARE tests with their original values.}
  \label{ex1_para}
\end{figure*}

\section{Flux Emergence Simulation and Data-driven MHD Model}
\label{sec:model}

Same as \citet{Toriumi2020}, we use the boundary data sliced from the
FES carried out by \citet{Toriumi2017}. The numerical code is based
on~\citet{Takasao2015}, and it solves the MHD process of an isolated,
twisted magnetic flux tube, initially placed in the convection zone,
buoyantly rising into the atmosphere. The full
set of basic resistive, adiabatic MHD equations in conservative form are solved by
a finite difference scheme with the spatial derivatives by the fourth-order central differences
and the temporal derivatives by the four-step Runge-Kutta scheme. Periodic
boundary conditions are applied for the $x$-direction and symmetric boundaries
for both the $y$- and $z$-directions. The initial background atmosphere
of the simulation is gravitationally stratified, which consists of an
adiabatically temperature-gradient convection zone ($z/H_0 < 0$, where
$H_0=170$~km is the length unit of the FES model), the cool isothermal
photosphere/chromosphere ($0 \leq z/H_0 < 18$), and the hot isothermal
corona ($z/H_0 \ge18$).  The flux tube is initially placed at
$z=-30 H_0$ with the form of
$B_x(r) = B_{\rm tube} \exp (-r^2/R_{\rm tube})$ and
$B_\phi (r) = qrB_x(r)$, where $r$ is the radial distance from the
tube's axis, $R_{\rm tube} = 3H_0$ the radius, $B_{\rm tube} = 7.5$~kG
the axial field strength, and $q=-0.2/H_0$ the twist intensity (the
negative sign indicates a left-handed twist). The middle of the tube
is made buoyant by a small density deficiency and the tube started to
emerge due to its own buoyancy.

As the time-dependent bottom boundary data for data-driven model, we
extracted the time sequence of 2D data at fixed heights from the FES
model. To test the behavior of the data-driven model with different
inputs, data sets at three different slices from the FES model are
used, including $z/H_0 = 0$ (referred to \texttt{Z00}, representing
the photosphere), $z/H_0 = 10$ (\texttt{Z10}, i.e., the chromospheric
layer) and $z/H_0 = 20$ (\texttt{Z20}, i.e., the coronal base). The
slices were sampled at every $\Delta t/\tau_0 = 1$ (where $\tau_0 = 25$~s
is the time unit of the FES model) from $t/\tau_0 = $ 0 to 500. The
slices spanned over $(-165, -165) \le (x/H_0, y/H_0) \le (165, 165)$,
resolved by the uniform grid spacing of $200 \times 200$, thus the
resolution is $\Delta = 1.65 H_{0}$. The data sets include three
components of magnetic and velocity field, i.e.,
$B_x, B_y, B_z, v_x, v_y$, and $v_z$.

\begin{figure*}
  \centering
  \includegraphics[width=\textwidth]{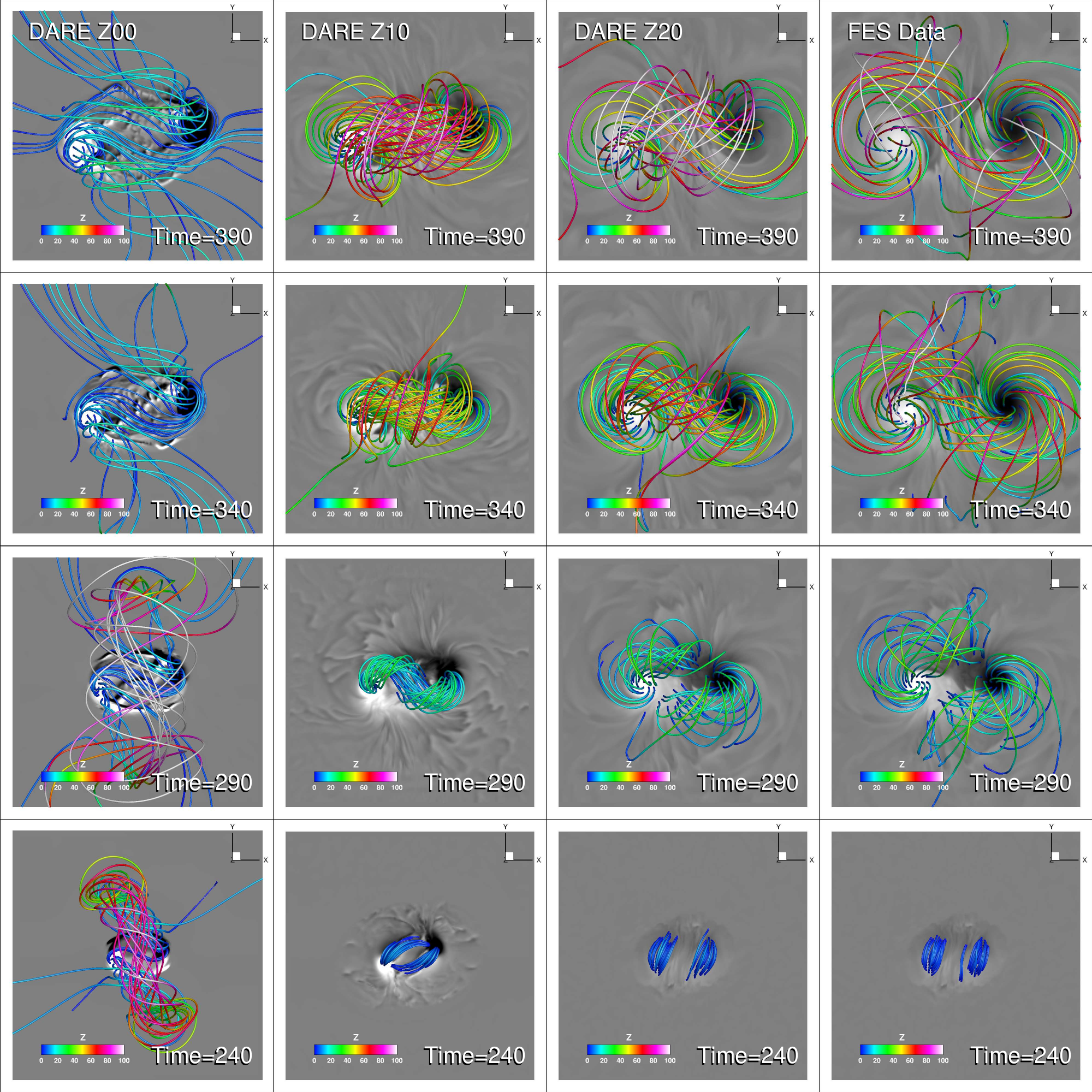}
  \caption{Comparison of magnetic field lines at four time snapshots
    in the three DARE tests with the original FES data.  Note that
    since different models have different bottom boundaries, the
    footpoints of the magnetic field lines are not generally idential,
    but they are all located within the main polarities. Colors of the
    field lines denote the height above the bottom surface.}
  \label{ex1_bline}
\end{figure*}

As we have mentioned in Section~\ref{sec:intro}, the Lorentz force in
the boundary data is an important factor affecting the output of the
data-driven model. Thus we quantitatively assess the Lorentz force as
well as torque in the data of different layers. The sum of Lorentz
force $\vec F$ (and torque $\vec T$) in a volume can be expressed by
surface integral~\citep{Aly1989, Sakurai1989}, and by neglecting
contribution from the side and top boundaries, they can be written as
integrals of the 2D data sets:
\begin{equation}\label{force}
\begin{split}
  F_x = -\frac{1}{4\pi}\int B_x B_z dxdy,  \\
  F_y =-\frac{1}{4\pi}\int B_y B_z dxdy,  \\
  F_z =-\frac{1}{8\pi}\int (B_z^2-B_x^2-B_y^2) dxdy,
  \end{split}
\end{equation}
and
\begin{equation}\label{torque}
\begin{split}
  T_x = -\frac{1}{8\pi} \int y(B_z^2-B_x^2-B_y^2) dxdy, \\
  T_y = \frac{1}{8\pi} \int x(B_z^2-B_x^2-B_y^2) dxdy, \\
  T_z = \frac{1}{4\pi} \int (yB_xB_z-xB_yB_z) dxdy.
\end{split}
\end{equation}
Following \citet{DuanA2020}, to assess the data with respect to the
force-free condition, the forces can be normalized by the integrated
magnetic pressure force $F_p$ that is given by
\begin{equation}
  F_p = \left |\int \left(-\nabla \frac{B^2}{8\pi}\right) dV \right |
  = \frac{1}{8\pi}\int (B_x^2+B_y^2+B_z^2) dxdy,
\end{equation}
and the normalized forces are ratios defined as $f_x = |{F_x}|/{F_p}$,
$f_y = |{F_y}|/{F_p}$, and $f_z = |{F_z}|/{F_p}$, respectively.
Similarly, the normalized torques are defined as
$t_x = |{T_x}|/{T_p}$, $t_y = |{T_y}|/{T_p}$, and
$t_z = |{T_z}|/{T_p}$, where $T_p$ is magnitude of the net torque
induced by only the magnetic pressure force,
$T_p = |\int \vec r \times (-\nabla \frac{B^2}{8\pi}) dV| =
\sqrt{T_{px}^2+T_{py}^2}$ where
\begin{equation}
\begin{split}
T_{px} = \frac{1}{8\pi} \int y(B_x^2+B_y^2+B_z^2) dxdy,\\
T_{py} = \frac{1}{8\pi} \int x(B_x^2+B_y^2+B_z^2) dxdy.
  \end{split}
\end{equation}
For a field being close to force-free, it must have all the ratios
much less than unity, i.e., $(f_x, f_y, f_z) \ll 1$ and
$(t_x, t_y, t_z) \ll 1$. \citet{Metcalf1995} suggested that the
magnetic field can be considered as force-free if the normalized
forces are all less than or equal to $0.1$, and this criterion is
widely accepted by the later studies~\citep{ moon_forcefreeness_2002,
  tiwari_force-free_2012, liu_statistical_2013, liu_research_2015,
  jiang_influence_2019, DuanA2020}. On the other hand, for a strongly
non force-free field, the ratios can be close to unity, meaning that
the magnetic pressure force and the tension force are so unbalanced
that the net Lorentz force is comparable to one of its components, the
total magnetic pressure force.

In \Fig~\ref{residual_force}, we show these metrics of the FES data.
As can be seen, the FES data at the photosphere (i.e., \texttt{Z00}
slice) has a very large verical Lorentz force and horizontal torque,
since $f_z$, $t_{x}$ and $t_{y}$ are close to $1$ for the whole
duration of evolution with the flux injection to its saturation. Note
that it is due to the perfect symmetry of two emerging polarities in
the simulations that the net forces in horizontal directions (i.e.,
$f_x$ and $f_y$) and the net torques in vertical direction ($t_z$) are
both nearly zero.  For comparison, \citet{DuanA2020} showed that these
parameters derived from \SDO/HMI data of emerging ARs are close to
$0.1$, which is much smaller than those of the simulated
data~\footnote{ It is possible that the effective height of magnetic
  observation in HMI is somewhat different from \texttt{Z00} of the
  FES model; see \citet{DuanA2020} for further details.}.  At the two
higher levels (i.e., \texttt{Z10} and \texttt{Z20}), the Lorentz force
and torque in the early phase of emergence are still large, but
decrease substantially with the increase of the magnetic flux,
indicating that at these levels the fields are relaxing to a more
force-free state. But it should be noted that they are still larger
than typical observed values~\citep[see also][]{DuanA2020}.

\begin{figure*}
  \centering
  \includegraphics[width=\textwidth]{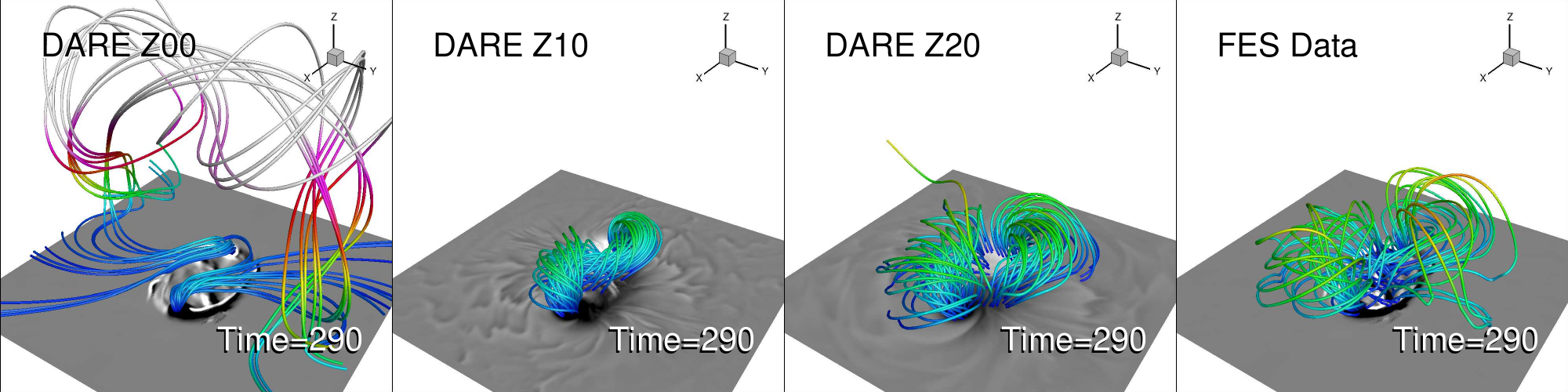}
  \includegraphics[width=\textwidth]{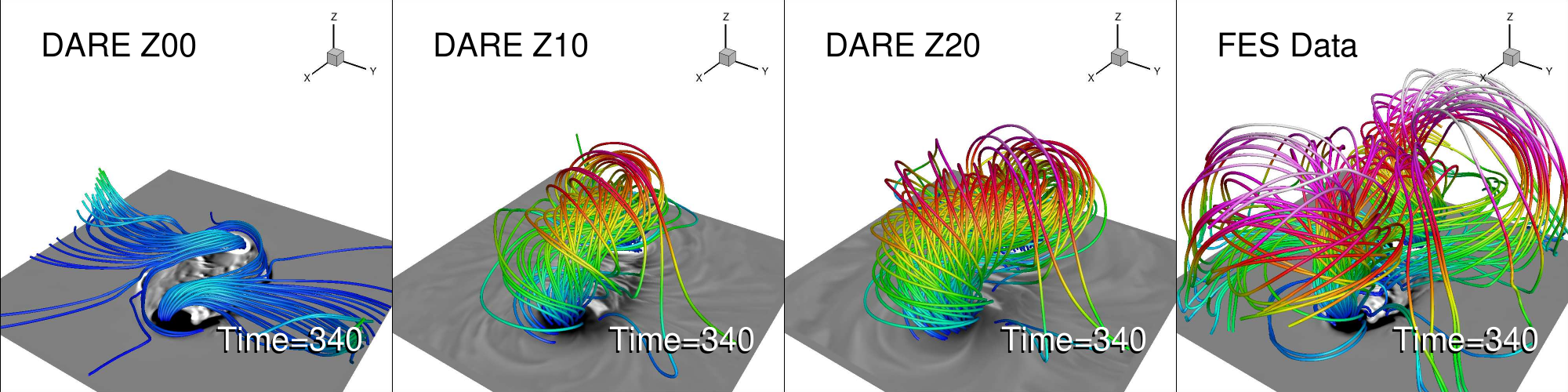}
  \caption{Comparison of magnetic field lines at $t=290$
    and $t=340$ in a 3D view.}
  \label{ex1_bline3d}
\end{figure*}

\begin{figure*}
  \centering
  \includegraphics[width=0.8\textwidth]{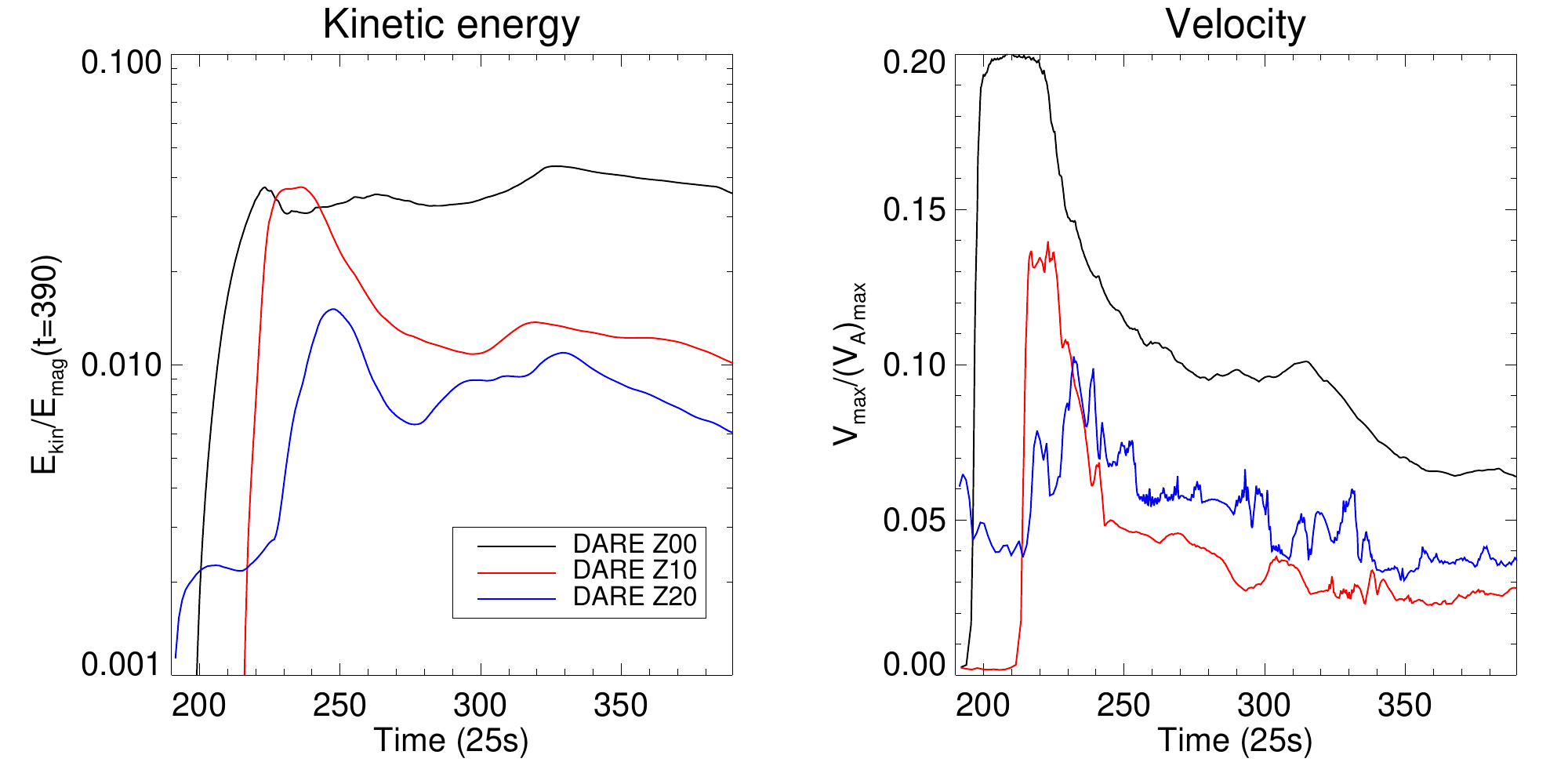}
  \caption{Comparison of kinetic energy and maximal velocity in the three tests.}
  \label{ex1_kinetic}
\end{figure*}

\citet{Jiang2016NC}'s data-driven active-region evolution (DARE) model
is aimed for reproducing the coronal magnetic field evolution by
assuming the bottom surface as the base of the corona. Like the FES
model, it solves the full set of the MHD equations with both gas
pressure and solar gravity. Also the adiabatic energy equation is used. The numerical code
that solves the MHD equations is developed based on the space-time conservative element and solution element (CESE) scheme~\citep{Jiang2010} with a second-order accuracy, and in this code, all MHD
variables are specified at the corners of the grid cells and no ghost cells (or guard cells) are used at the boundaries.
Since the data-driven simulation is started before the field emerges
into the photosphere, the initial conditions consist of a zero
magnetic field and a plasma in hydrostatic equilibrium. In all experiments,
we used the same initial background atmosphere: plasma stratified by
the solar gravity with a uniform temperature of typical coronal value
($T=10^6$~K). The bottom boundary of the model is placed at the
coronal base with a fixed number density of $10^9$~cm$^{-3}$ (also a
typical value in the corona). After non-dimensionalization, the plasma
density and gas pressure at the bottom boundary are both $1$. For all
experiments, we used the same computational grid, which has an extent
of
$(-192, -192, 0) \le (x/\Delta, y/\Delta, z/\Delta) \le (192, 192,
384)$
with grid resolution of $\Delta$, where $\Delta = 1.65 H_{0}$ is the
resolution of the input FES data slice. Note that the computational
box is nearly twice the size of the FES magnetogram, thus it can
effectively reduce the influence from the side and top boundaries
(where the plasma variables are all fixed, the horizontal magnetic
field is linearly extrapolated and the normal magnetic field is
modified to fulfill the divergence-free constraint). On the bottom
boundary, the temporally-evolving magnetic field and velocity at a
given slice from the FES model are introduced sequentially while the
plasma density and temperature are fixed.
The boundary data of the FES model is provided at a fixed time cadence of 25 s. However, because the time step of the DARE model is variable according to the CFL stable condition, we use a linear interpolation (in the time domain) at each DARE time step to obtain the boundary data at the exact moment. The magnetic field and
velocity outside the field-of-view of the input FES slice are assumed
to be zero. For saving computing time, we run all the data-driven
simulation with time from $t/\tau_{0}=190$ to $390$, since before
$t/\tau_{0}=190$ the magnetic flux has not yet reached the photosphere
$z/H_{0}=0$.

\section{Results}
\label{sec:res}


We test the behavior of the DARE model by driving it with different
slices of FES data. Since the magnetic field strength of different
slices are different, and to make the comparison within the very same
conditions, the magnetic field of each slice are scaled such that the
maximum of $B_z$ at the bottom boundary is $10$ (non-dimensionalized)
in the whole evolution. Thus the maximum Alfv{\'e}n speed is
approximately $10$ (normalized by the sound speed), and smallest
plasma $\beta$ is $0.02$, representing a typical low-$\beta$
environment in the corona.

We compare the DARE results with the original FES 3D data in exactly
the same volume of
$(-100, -100, z_{\rm FES}) \le (x/\Delta, y/\Delta, z/\Delta) \le
(100, 100, z_{\rm FES} + 200)$
where $z_{\rm FES} =$ \texttt{Z00}, \texttt{Z10}, and \texttt{Z20},
respectively, and the same resolution of $dx = dy = dz = \Delta$. Two
global quantities of the magnetic field are calculated, which are the
magnetic energy and relative magnetic helicity. Here the relative
helicity is obtained by using the method of
\citet{Valori2012}. \Fig~\ref{ex1_para} shows the results of magnetic
energy and helicity. It can be seen that the DARE-\texttt{Z00} results are
larger than that of the FES by an order of magnitude in both
magnetic energy and helicity, in agreement with the results shown
by~\citet{Toriumi2020}. The DARE-\texttt{Z10} and DARE-\texttt{Z20}
results appear to be increasingly better, in particular, the
DARE-\texttt{Z20} produces both energy and helicity matching their
original values very closely.

\Figs~\ref{ex1_bline} and \ref{ex1_bline3d} present the magnetic field
lines. The DARE-\texttt{Z00} shows a drastic expansion at the very
early phase like the previous result shown by~\citet{Toriumi2020},
which makes the magnetic field lines mostly extend out of the box
after around $t=300$. Such a drastic expansion is not seen in the
results of DARE-\texttt{Z10} and \texttt{Z20}, both of which consist
of a relatively coherent flux rope at $t=340$. As can be seen from the
field lines, the DARE results are getting better agreement with the FES
with the increasing geometrical level of input data.

Finally, \Fig~\ref{ex1_kinetic} shows the evolutions of the ratio of
the kinetic energy (and the largest velocity) to the magnetic energy
(and the largest Alfv{\'e}n speed) in the three experiments. For
DARE-\texttt{Z00}, the kinetic energy keeps at the level of 4\% of the
total magnetic energy over the whole computational duration. This is
because the large Lorentz force at the boundary continuously drives
the plasma flow. For DARE-\texttt{Z10}, the kinetic energy first rises
to the same value as in DARE-\texttt{Z00}, but drops very quickly to
roughly 1\% of magnetic energy, as the Lorentz force at the bottom
boundary deceases.  As the \texttt{Z20} data is even closer to
force-free, the kinetic energy is smaller. The similar conclusions can
be seen from the ratio of the flow speed to the Alfv{\'e}n speed. For
instance, the \texttt{Z00} data drives a velocity reaching $0.2$ of
the largest Alfv{\'e}n speed, indicating that the strong Lorentz force
results in a very dynamic flow. For the other two experiments, the
velocity is smaller than the Alfv{\'e}n speed by approximately two
orders of magnitude, suggesting that the MHD system evolves in a
quasi-static way.

\section{Conclusions}
\label{sec:con}
\citet{Toriumi2020} have investigated different types of data-driven
coronal field models using an FES as a reference or ground truth data
set. It is found that in the FES the photospheric magnetic field
contains a strong Lorentz force, which was too large for the DARE
model to reasonably reproduce
the coronal magnetic field because unphysical, spurious flows are excited.

In this paper, we have comprehensively compared the DARE simulations
driven by three different levels of data from the FES, corresponding
to the photosphere (\texttt{Z00}), the chromosphere (\texttt{Z10}) and
the base of corona (\texttt{Z20}). The key difference of the three
sets of data is the extents of the Lorentz force. In the \texttt{Z00}
data, the field is far from a force-free state with a normalized
Lorentz force and torque approaching unity. In the \texttt{Z10} and
\texttt{Z20}, the normalized force and torque decrease
substantially. As a result, the DARE model attains much better results
of the coronal magnetic field by using the \texttt{Z10} and
\texttt{Z20} data than using the \texttt{Z00} data. In particular, by
using the \texttt{Z20} data, which has the smallest Lorentz force
among all, the DARE model reproduced the magnetic energy and helicity
in a high confidence. Thus, in this experiment, we have confirmed that
the Lorentz force in the boundary data is a key issue influences the
results of the DARE model. It would be better to use observations for
the chromosphere, or even higher in the coronal base as the driving
boundary data for the DARE model since the field at the higher levels
would be more force-free.  Nevertheless, as the recent statistical
study by \citet{DuanA2020} reveals, the photospheric field of
dynamically emerging flux is in fact close to the force-free state
unlike the FES model (in particular in \texttt{Z00}).  Therefore, it
is still possible to use the observed photospheric field as an
approximate of coronal base field for driving the data-driven models,
such as the DARE. Future investigations using realistic convective
FES models that yield more relaxed emergence~\citep[e.g.,][]{Toriumi2019S, Hotta2020} may allow us to \textbf{understand how accurate the data-driven models actually are.}


\acknowledgments C.J. acknowledges support by National Natural Science
Foundation of China (41822404, 41731067), the Fundamental Research
Funds for the Central Universities (Grant No.HIT.BRETIV.201901), and
Shenzhen Technology Project JCYJ20190806142609035. S.T. was supported
by JSPS KAKENHI Grant Numbers JP15H05814 (PI: K. Ichimoto) and
JP18H05234 (PI: Y. Katsukawa), and by the NINS program for
cross-disciplinary study (Grant Numbers 01321802 and 01311904) on
Turbulence, Transport, and Heating Dynamics in Laboratory and
Solar/Astrophysical Plasmas: ``SoLaBo-X''. The computational work of
the DARE model was carried out on TianHe-1(A), National Supercomputer
Center in Tianjin, China.  Numerical computations of the FES model
were carried out on Cray XC50 at Center for Computational
Astrophysics, National Astronomical Observatory of Japan.


\end{document}